\documentclass[onecolumn]{aa}

\bibpunct{(}{)}{;}{a}{}{,}

\usepackage{graphicx}
\usepackage[varg]{txfonts}
\newcommand{\hii}{H\textsc{ii}}
\def\ks{km s$^{-1}$}

\def\m{$^\prime$}
\def\s{$^{\prime\prime}$}

\def\cm3{cm$^{-3}$}

\def\2{$^{12}$CO}
\def\3{$^{13}$CO}
\def\msol{M$_\odot$}

\def\cm2{cm$^{-2}$}
\begin{document}

\title{The southern molecular environment of SNR G18.8+0.3}
\author {S. Paron \inst{1,2,3}
\and M. Celis Pe\~{n}a \inst{1}
\and M. E. Ortega \inst{1}
\and A. Petriella \inst{1,3}
\and M. Rubio \inst{4}
\and G. Dubner \inst{1}
\and E. Giacani \inst{1,2}
}

\institute{Instituto de Astronom\'\i a y F\'\i sica del Espacio (IAFE),
             CC 67, Suc. 28, 1428 Buenos Aires, Argentina\\
             \email{sparon@iafe.uba.ar}
\and FADU - Universidad de Buenos Aires, Ciudad Universitaria, Buenos Aires
\and CBC - Universidad de Buenos Aires, Ciudad Universitaria, Buenos Aires
\and Departamento de Astronom\'{\i}a, Universidad de Chile, Casilla 36-D, Santiago, Chile}

\offprints{S. Paron}

   \date{Received <date>; Accepted <date>}

\abstract{}{In a previous paper we have investigated the molecular environment towards the eastern border of the SNR G18.8+0.3. 
Continuing with the study of the surroundings of this SNR, in this work we focus on its southern border, which in the radio
continuum emission shows a very peculiar morphology with a corrugated corner and a very flattened southern flank.}
{We observed two regions towards the south of SNR G18.8+0.3 using the Atacama
Submillimeter Telescope Experiment (ASTE) in the \2 J=3--2. One of these regions was also 
surveyed in \3 and C$^{18}$O J=3--2. The angular and spectral resolution of these observations were 22\s, and 0.11 \ks. 
We compared the CO emission to 20 cm radio continuum maps obtain as part of the Multi-Array Galactic Plane Imaging Survey
(MAGPIS) and 870 $\mu$m dust emission extracted from the APEX Telescope Large Area Survey of the Galaxy.}
%Our observations were complemented with data of the continuum emission at 870 $\mu$m extracted from the ATLASGAL.}
{We discovered a molecular feature
with a good morphological correspondence with the SNR's southernmost corner. 
In particular, there are
indentations in the radio continuum map that are complemented by protrusions in the
molecular CO image, strongly suggesting that the SNR shock is interacting with a
molecular cloud.
%In particular, the indentations in the morphology
%of the SNR radio continuum emission are complemented by protrusions in the molecular cloud, strongly suggesting that the 
%shock front is interacting with the cloud. 
Towards this region we found that 
the \2 peak is not correlated with the observed \3 peaks, which are likely related to a nearby \hii~region. 
Regarding the most flattened border of SNR G18.8+0.3, 
where an
interaction of the SNR with dense material was previously suggested, our \2 J=3--2 
map show no obvious indication that this is occurring.
%where it was suggested an interaction with dense material, strikingly 
%our \2 J=3--2 observations do not show any conspicuous correspondence with the SNR.
}{}

\titlerunning{The southern environment of SNR G18.8+0.3}
\authorrunning{S. Paron et al.}


\maketitle

\section{Introduction}

The investigation of the surroundings of supernova remnants (SNRs) can provide useful information for a wide range of 
research fields, as the physical and chemical changes induced by the passage of a shock wave, dust production and destruction,
the possibility of triggering stellar formation, production of high-energy gamma rays, etc. Multiwavelength studies 
are required to explore these effects. 

\begin{figure}[h]
\centering
\includegraphics[width=8.9cm]{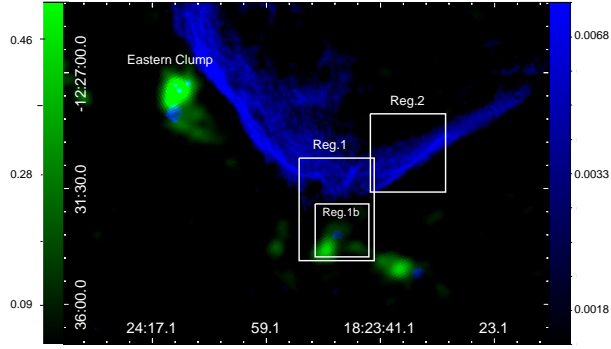}
\caption{Two-color image of the SNR G18.8+0.3 and its surroundings. The radio continuum emission at 20 cm (from MAGPIS) is displayed in blue 
(right color bar), and the ATLASGAL emission at 870 $\mu$m in green (left color bar). Both color ranges are in Jy beam$^{-1}$. 
The surveyed regions in the molecular lines using ASTE are indicated.}
\label{present}
\end{figure}

The SNR G18.8+0.3, with a peculiar shape with flattened borders along the east and south, has been suggested to be interacting 
with dense molecular gas towards such directions \citep{dubner99,dubner04,tian07}. \citet{paron12}, hereafter Paper\,I, 
investigated the molecular gas towards the eastern border of the SNR G18.8+0.3. 
They found a dense molecular clump, close to the shock front of the SNR but not in contact with it, containing
a complex of \hii~regions and massive young stellar objects (YSOs) deeply embedded. This region is indicated in Fig.\,\ref{present} 
as `Eastern Clump'.
Figure\,\ref{present} shows the SNR G18.8+0.3 in radio continuum at 20 cm as extracted from 
the MAGPIS \citep{helfand06} (in blue) and its surroundings as seen at the 870 $\mu$m emission extracted
from the ATLASGAL \citep{beuther12} (in green). 
Southwards of the SNR, where the molecular cloud appears to be in contact with SNR shock front (see Figure\,1 in Paper\,I), there are 
a few \hii~regions and submillimeter dust compact sources, making this region an interesting target to study the interplay
between the SNR shock front, surrounding molecular gas, and \hii~regions. 

Using molecular lines we investigate two regions towards the south of SNR G18.8+0.3; Reg.1 and the subregion Reg.1b, 
were selected to investigate the possible causes for the peculiar morphology of 
the SNR southernmost `corner', and the molecular gas towards the \hii~region G018.630+0.309 \citep{anderson11} and the 
ATLASGAL source 018.626+00.297 \citep{urqu14}, while
Reg.2 was chosen to investigate the molecular environment towards the most flattened border of the SNR
where originally it was suggested an interaction with a dense cloud \citep{dubner99}. 
As in Paper\,I, here we assume a distance of $14\pm1$ kpc for SNR G18.8+0.3.
Concerning to the \hii~region G018.630+0.309, \citet{anderson12} could not resolve the distance ambiguity between 1.5 and 14.7 kpc.

\section{Observations}

The observations of the molecular lines were carried out between September 7 and 13, 2013
with the 10 m Atacama Submillimeter
Telescope Experiment (ASTE; \citealt{ezawa04}). We used the CATS345 receiver, which is a two-single
band SIS receiver remotely tunable in the LO frequency range of 324-372 GHz. 
The surveyed regions are shown in Fig.\,\ref{present} and summarized in Table\,\ref{surveyed}. The observations were 
performed in position switching mode and the mapping grid spacing was 20\s~in all cases. The \2 J=3--2 line (345.796 GHz)  
was observed towards Regs.\,1, 2. In Reg.1b we also observed C$^{18}$O J=3--2 (329.330 GHz), and \3 J=3--2 (330.587 GHz).
The integration time was 20 sec for the \2 and 50 sec for the other molecular lines in each pointing.

\begin{table}
\caption{Surveyed regions using ASTE.}
\label{surveyed}
\centering
\begin{tabular}{lcc}
\hline
\hline
Region            & Center (R.A., dec.) & Size \\
\hline
Reg.1          &   18:23:48.0, $-$12:32:14.5 &   3\m$\times$4\m     \\
Reg.1b         &   18:23:47.1, $-$12:33:12.1 &   2\m$\times$2\m      \\
Reg.2          &   18:23:38.0, $-$12:30:04.6 &   3\m$\times$3\m    \\
\hline
\end{tabular}
\end{table}

We used the XF digital spectrometer with bandwidth and spectral resolution set to
128 MHz and 125 kHz, respectively.
The velocity resolution was 0.11 \ks, the half-power beamwidth (HPBW) 22\s,
and the main beam efficiency was $\eta_{\rm mb} \sim 0.65$.
The spectra were Hanning smoothed to improve the signal-to-noise ratio and only linear or some second order
polynomia were used for baseline fitting.
The data were reduced with NEWSTAR\footnote{Reduction software based on AIPS developed at NRAO,
extended to treat single dish data
with a graphical user interface (GUI).} and the spectra processed using the XSpec software
package\footnote{XSpec is a spectral line reduction package for astronomy which has been
developed by Per Bergman at Onsala Space Observatory.}.

\section{Results and discussion}

\subsection{Regions 1 and 1b}

\begin{figure}[h]
\centering
\includegraphics[width=8cm]{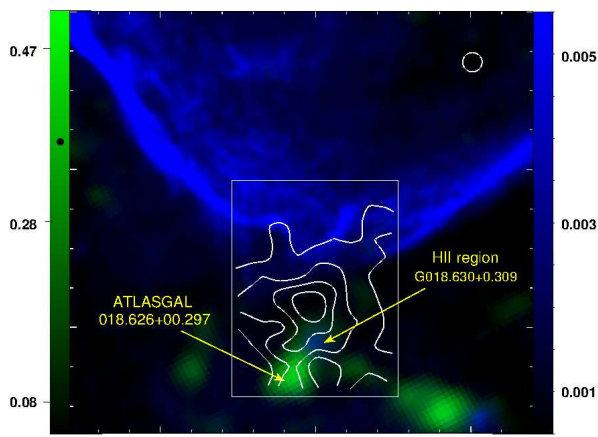}
\caption{Two-color image towards Reg.1 with the radio continuum emission at 20 cm (from MAGPIS) displayed in blue (right color bar),
and the ATLASGAL emission at 870 $\mu$m in green (left color bar). Both color ranges are in Jy beam$^{-1}$. 
The \2 J=3--2 emission integrated between
5 and 30 \ks~is presented in contours, with levels of 32, 42, 56, and 68 K \ks. The beam of the molecular observations is shown at the top right corner. The positions of the \hii~region G018.630+0.309 and the compact dust source
ATLASGAL 018.626+00.297 are indicated. }
\label{12coReg1}
\end{figure}

In Reg.1, \2 J=3--2 emission is detected at velocities between 5 and 30 \ks, very
much like what was observed in the eastern region (Paper\,I). The emission integrated over the mentioned velocity range 
is displayed as contours over a two-color image showing the 20 cm and 870 $\mu$m continuum images in Fig.\,\ref{12coReg1}.
%In Reg.1, the \2 J=3--2 emission is detected between 5 and 30 \ks, similarly to what was observed in the eastern region reported
%in Paper\,I. Figure\,\ref{12coReg1} displays in contours the \2 emission integrated along the mentioned velocity range over a two-color 
%image displaying the 20 cm and 870 $\mu$m continuum emissions. 
%The positions of the \hii~region G018.630+0.309 and the compact dust source 
%ATLASGAL 018.626+00.297 are indicated. 
Importantly, the boundaries of the radio continuum emission from the SNR and the \2 emission from the 
molecular cloud show a number of common features, strongly suggesting that the shock front is interacting with the cloud. In particular,
the most noticeable indentations in the border of the SNR are complemented by protrusions in the molecular cloud. 
In addition, towards the south the contour at 56 K \ks~of the \2 emission has a peculiar curvature apparently in coincidence with
the \hii~region G018.630+0.309, suggesting that this object is also associated with the molecular structure.

\begin{figure}[h]
\centering
\includegraphics[width=6cm]{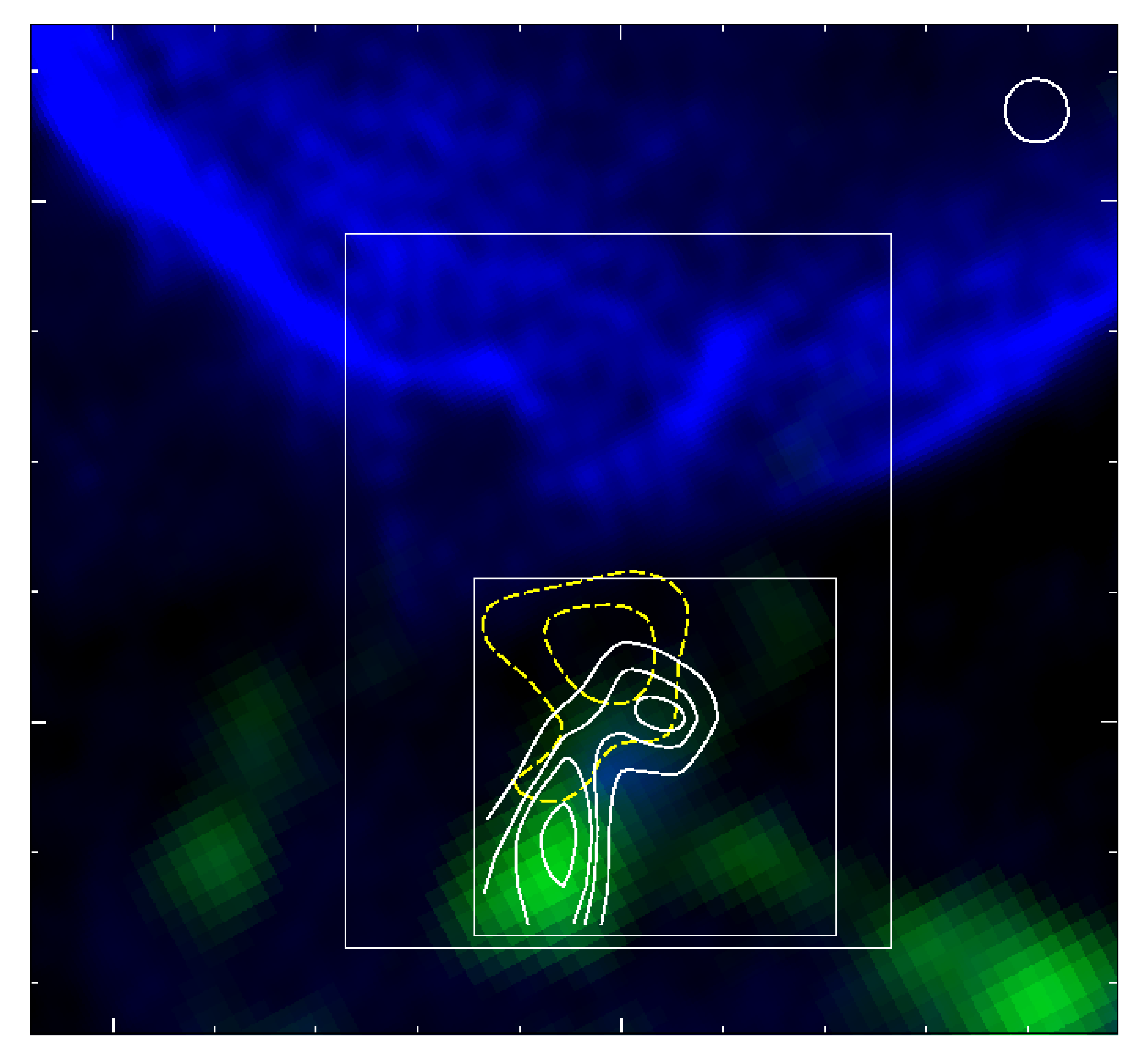}
\caption{Two-color image towards Reg.1b with the radio continuum emission at 20 cm (from MAGPIS) displayed in blue,
and the ATLASGAL emission at 870 $\mu$m in green. The color ranges are the same as in Fig.\,\ref{12coReg1}. 
For context, the box of Reg.1 and the 56, and 68 K \ks~contours 
(in dashed yellow line) of the \2 emission presented in Fig.\,\ref{12coReg1} are also shown. 
Contours of the \3 J=3--2 emission integrated between 15 and 26 \ks~are shown in white, with levels of 12.0, 15.0, 16.5, and 20.0 K \ks. 
The beam of the molecular observations is shown at the top right corner.}
\label{13coReg1b}
\end{figure}

Figure\,\ref{13coReg1b} displays the \3 J=3--2 emission integrated between 15 and 26 \ks~towards Reg.1b in white contours. 
For context, the 56, and 68 K \ks~contours of the \2 emission shown in Fig.\,\ref{12coReg1} are also shown in Fig.\,\ref{13coReg1b}
(yellow dashed contours). 
The \3 emission reveals a molecular feature, composed by two clumps, that partially surrounds the \hii~region G018.630+0.309. 
One of these clumps, the
southernmost, exactly overlaps the source ATLASGAL 018.626+00.297. Curiously the most intense emission of the \2  
is not correlated with the \3 peaks. Concerning the C$^{18}$O J=3--2 line, we only detect emission (a single spectrum) 
over the maximum of the ATLASGAL source, implying that this is the region with 
highest density.

%\begin{figure}[h]
%\centering
%\includegraphics[width=6.5cm]{FigSpectC18O.ps}
%\caption{C$^{18}$O J=3--2 spectrum obtained towards the source ATLASGAL 018.626+00.297 (at 18:23:49.5, $-12$:33:52, J2000). 
%The rms noise level is 0.22 K.}
%\label{c18o}
%\end{figure}

In order to discuss the molecular gas in Reg.1 and Reg.1b we denote the structure delimited by the 56 K \ks~contour 
(the dashed yellow contour in Fig.\,\ref{13coReg1b})  as the ``\2-clump'', the feature delimited by a circle of 25\s~in radius centered 
at the northern maximum in the \3 emission is designated as the ``\3-clumpN'', and the southern \3 clump delimited by the  16.5 K \ks~contour 
as the  ``\3-clumpS'' (see  Fig\,\ref{13coReg1b}). 
We estimated the mass of these features applying a variety of methods, namely: by assuming local thermodynamic 
equilibrium (LTE), from the CO luminosity (in the case of the \2-clump), from the thermal dust emission at 870 $\mu$m (in the
case of the \3-clumpS), and based on the virial theorem. To obtain the LTE mass, we follow
the procedures described in Paper\,I to estimate the column densities of N(\2) and N(\3). Table\,\ref{lteparam} presents the derived and used
parameters to perform that. The column density values represent the total column density as obtained from summation over all beam positions 
belonging to each molecular structure described above. To obtain the mass from the CO luminosity we use (e.g. \citealt{bert93}):
$ {\rm N(H_{2})} = 2 \times 10^{20}~{\rm W(^{12}CO)}$
to estimate N(H$_{2}$), where W(\2) is the velocity integrated \2 emission along the whole \2-clump structure. 
Finally, once obtained the total N(H$_{2}$), either from the LTE approximation or from
the CO luminosity, the mass is derived using ${\rm M} = \mu~m_{{\rm H}}~d^{2}~\Omega~{\rm N(H_{2}})$, 
where $\Omega$ is the solid angle subtended by the beam size, $m_{\rm H}$ is the hydrogen mass,
$\mu$, the mean molecular weight, assumed to be 2.8 by taking into account a relative helium abundance
of 25 \%, and $d$ is the distance. 

Assuming that the \3-clumpS is related to 
the ATLASGAL source 018.626+00.297, we estimate its mass from the thermal dust emission at 870 $\mu$m following
the procedure presented in \citet{csengeri14}. We use an integrated flux of S$_{\rm int} = 3.31$ Jy \citep{urqu14},
and assume a typical dust temperature of T$_{\rm d} = 15$ K.

Finally, the virial mass of the three molecular structures is derived from the relation:
\begin{equation}
{\rm M_{vir}} = \frac{5~{\rm R_{clump}} (\Delta{\rm v})^{2}}{8~{\rm G}~{\rm ln(2)}} 
\label{virial}
\end{equation}
where G, R$_{\rm clump}$ and $\Delta$v are the gravitational constant, the clump radius (assuming
a spherical geometry), and the line width, respectively. Following \citet{saito07} we define the clump radius as
the deconvolved radius calculated from ${\rm R_{clump}} = \sqrt{\frac{S - {\rm beam~area}}{\pi}}$,
where $S$ is the area inside the clump. 
The velocity width $\Delta$v was derived through a Gaussian fitting to the averaged spectrum obtained for the molecular features.
The deconvolved radius (${\rm R_{clump}}$) are 2.95, 1.52, and 1.05 pc, and the $\Delta$v $=$ 7.05, 3.80, and 5.10 \ks~for the \2-clump,
\3-clumpN, and \3-clumpS, respectively. 
All mass results are presented in Table\,\ref{mass}.
 
\begin{table}
\caption{Used and derived parameters for the LTE mass.}
\label{lteparam}
\small
\centering
\begin{tabular}{lccc}
\hline
\hline
Parameter                & \2-clump              & \3-clumpN              & \3-clumpS                \\
\hline
T$_{\rm ex}$ (K)         & 16.6                    &  12.5                    &  11.0                       \\
$\tau^{12}$              & 15                    &    --                  &    --                     \\
$\tau^{13}$              & --                    &   1.3                  &    3.5                    \\
N($^{12}$CO) (cm$^{-2}$)\tablefootmark{*}        & 7.1 $\times 10^{18}$   &    --                  &    --                   \\
N($^{13}$CO) (cm$^{-2}$)\tablefootmark{\dag}     & --                     &  1.5 $\times 10^{17}$  &    3.5 $\times 10^{17}$  \\
N(H$_{2}$) (cm$^{-2}$)   &  7.1 $\times 10^{22}$ & 7.5 $\times 10^{22}$   &  1.7 $\times 10^{23}$  \\
\hline
\end{tabular}
\tablefoot{
\tablefoottext{*}{To obtain the N(H$_{2}$) from the N($^{12}$CO) we assume the canonical abundance ratio [\2/H$_{2}$] $= 10^{-4}$. }
\tablefoottext{\dag}{To obtain the N(H$_{2}$) from the N($^{13}$CO) we assume [\3/H$_{2}$] $= 2 \times 10^{-6}$ (e.g. \citealt{simon01}).}
}
\end{table}

%\begin{table}
%\caption{Parameters used in the virial mass calculation.}
%\label{vir}
%\centering
%\begin{tabular}{lcc}
%\hline
%\hline
%Clump       &  R$_{\rm clump}$          &  $\Delta$v    \\
%            &   (pc)               &   (\ks)             \\
%\hline
%\2-clump    &  2.95    &  7.05     \\
%\3-clumpN   &  1.52    &  3.80     \\
%\3-clumpS   &  1.05    &  5.10     \\
%\hline
%\end{tabular}
%\end{table}

\begin{table}
\caption{Mass values.}
\label{mass}
\centering
\begin{tabular}{lcccc}
\hline
\hline
Clump       &  M$_{\rm LTE}$          &  M$_{\rm CO~lum.}$  & M$_{\rm dust}$   & M$_{\rm vir}$     \\
            &   (\msol)               &   (\msol)            &   (\msol)       &    (\msol)     \\     
\hline
\2-clump    &  $2.7 \times 10^{3}$    &  $6.0 \times 10^{3}$ & --                  &  $3.1 \times 10^{4}$   \\
\3-clumpN   &  $3.0 \times 10^{3}$    & --                   & --                  &  $4.5 \times 10^{3}$ \\
\3-clumpS   &  $7.0 \times 10^{3}$    & --                   & $5.5 \times 10^{3}$ &  $5.7  \times 10^{3}$\\
\hline
\end{tabular}
\end{table}

The only case in which the virial mass differs considerably from the mass
obtained by other methods is for the \2-clump. For this feature, the virial mass
is almost an order of magnitude larger than the LTE and CO luminosity based-estimates of the
mass. This suggests that this feature is not gravitationally bound.
%It can be appreciated that the only case in which the virial mass considerably differ from the mass 
%value obtained through other methods is in the \2-clump. 
%As presented in Table\,\ref{mass},  M$_{\rm vir}$ is almost one order of magnitude larger than the M$_{\rm LTE}$ and M$_{\rm CO~lum.}$, which can 
%indicate that the clump is not gravitationally bound. 

We assume as a working hypothesis that the \2-clump is a small molecular feature supported by external pressure.
The external pressure ($P_{ext}$) required for the \2-clump to remain bound, can be roughly evaluated from the following 
expression of the virial theorem (e.g. \citealt{kawamura98}): 
\begin{equation}
{2 U + \Omega - 4 \pi {\rm R_{clump}}^{3} P_{ext} = 0} 
\label{virialPext}
\end{equation}
where 
\begin{equation}
U = 3/2~{\rm M}~\frac{\Delta {\rm v}^{2}}{8 {\rm ln}(2)},~~ {\rm and}~~ \Omega = - \frac{3 {\rm G} {\rm M}^{2}}{5 {\rm R_{clump}}}
\label{virialPext2}
\end{equation}
Using a mean value between M$_{\rm LTE}$ and M$_{\rm CO~lum.}$ for the mass, we obtain that it is required a $P_{ext}/k_{B} \sim 1.4 \times 10^{6}$ 
cm$^{-3}$ K, with $k_{B}$ the Boltzmann constant, to keep the \2-clump bound. This value is at least one order of magnitude larger 
than the external pressures calculated in \citet{kawamura98} for a sample of clouds, where it is considered that the pressure is produced by 
the surrounding diffuse ambient gas. Thus, if this clump is supported by external pressure it is required a pressure source, which in 
the present context it is likely to be the SNR shocks running into the outer layers of the molecular gas. 
The derived $P_{ext}/k_{B}$ is very similar to the pressure
obtained towards a region associated with the face-on shock in the Vela SNR using FUSE observations \citep{sankrit01}, providing 
further support to our hypothesis that a SNR shock front contributes to maintain bound 
the observed \2-clump.

Finally we perform a SED analysis to the IR counterpart of \hii~region G018.630+0.30, the WISE source J182348.08-123316.1 \citep{cutri13,wri10}.
The fitting of the SED was done using the tool developed by
\citet{rob07}\footnote{http://caravan.astro.wisc.edu/protostars/} considering the fluxes at the WISE 3.4, 4.6, 12, and 22~$\mu$m
bands, an $A_v$ between
14 and 50 mag. (see Paper\,I) and a distance range of 13--15 kpc.
The SED results show that the source is indeed a young massive star (age $\sim$10$^{5}$ yrs and mass $\sim12$ \msol), suggesting
that its formation should be approximately coeval with the SN explosion.
If the SED is performed using the near distance (a range 1--3 kpc),
we find that the source should have a lower mass (about 3 \msol),
which would be unable to ionize the gas and generate an \hii~region. This is an indirect confirmation
that the \hii~region are located at the farther distance, suggesting that it should be embedded in the molecular
cloud southwards the SNR.

\subsection{Region 2}

\begin{figure}[h]
\centering
\includegraphics[width=5cm]{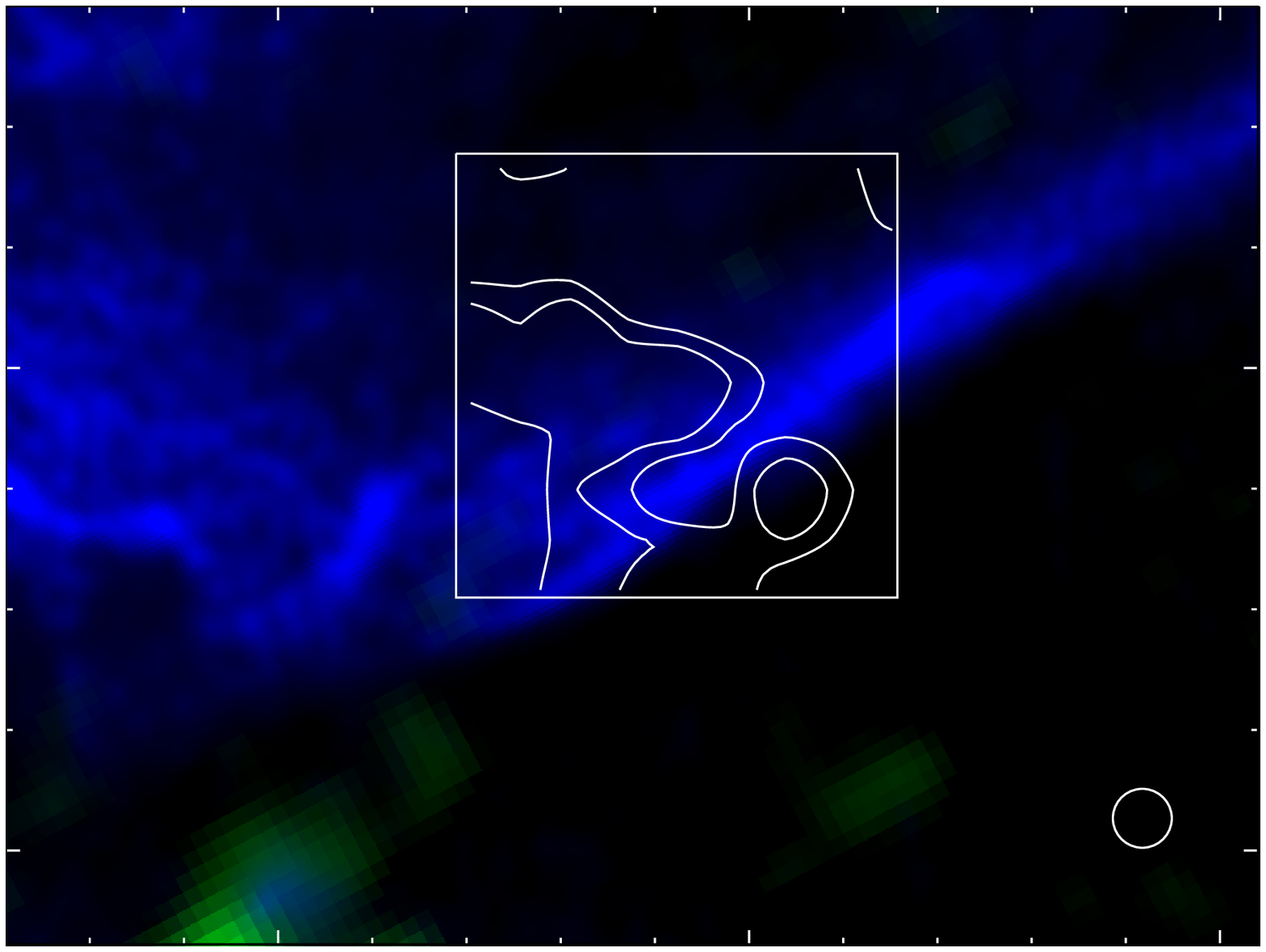}
\includegraphics[width=8cm]{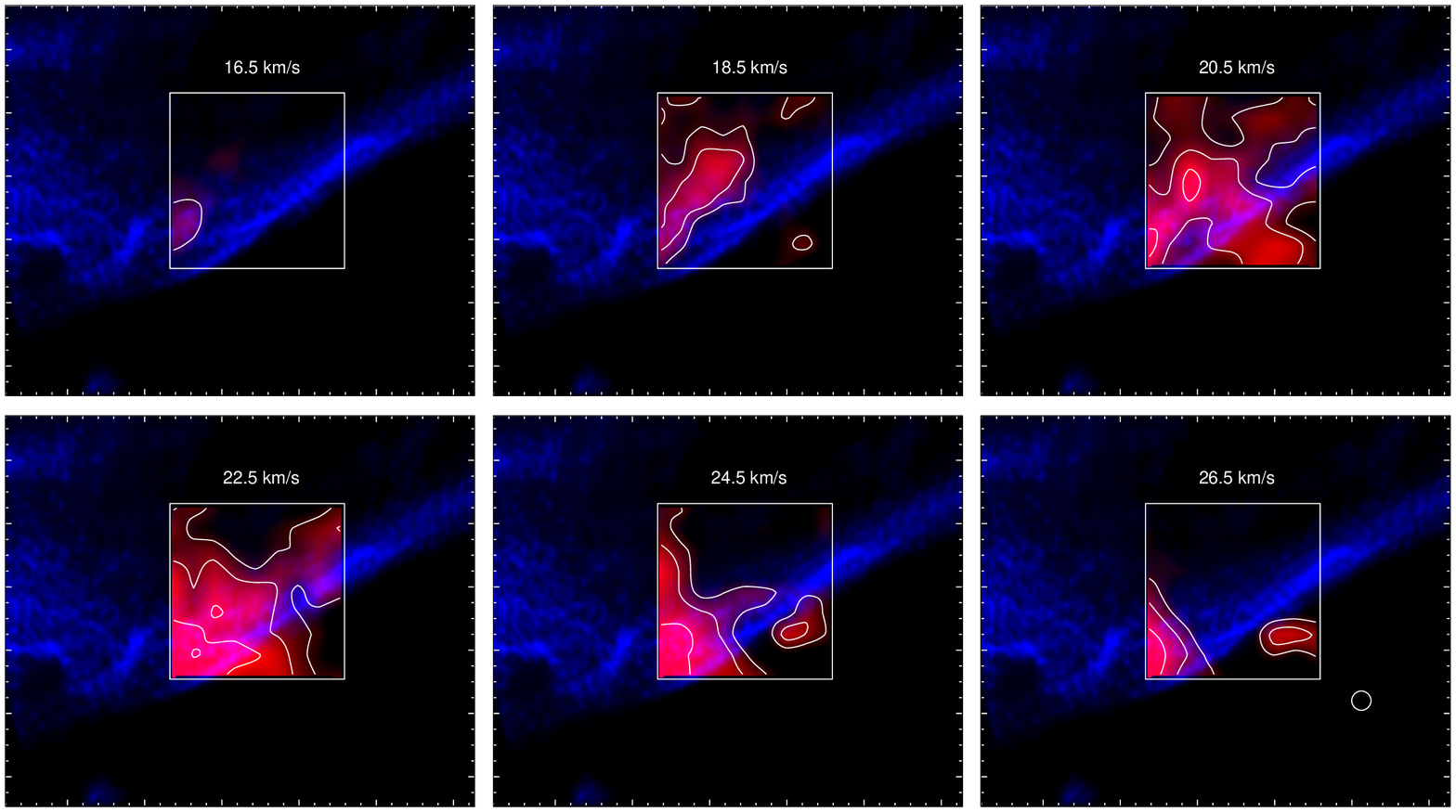}
\caption{Left: \2 J=3--2 emission integrated between 15 and 30 \ks, with levels of 14, 18, and 28 K \ks~(white contours) towards
Reg.2. The color ranges are the same as in Fig.\,\ref{12coReg1}. Right: channel maps of the \2 J=3--2 emission 
(red with white contours) integrated in steps of 
2 \ks. The contours levels are 2.5, 4.0, and 6.0 K \ks. 
%The indicated velocity in each panel corresponds to
%the first channel of the integrated interval.  
}
\label{12coReg2}
\end{figure}

The \2 emission from this region is significantly weaker than that detected
towards Reg.1.
By integrating the \2 J=3--2 emission between 15 and 30 \ks~(see Fig.\,\ref{12coReg2}-up), the whole velocity range where there is emission, 
we do not find any conspicuous morphological correspondence between the molecular gas and the most flattened border of the SNR which 
might suggests some kind of interaction. To appreciate in more detail the behavior of the molecular gas in this region, we present in 
Fig.\,\ref{12coReg2}-bottom the \2 emission in a series of channel maps integrated
in steps of 2 \ks. It can be seen that the molecular gas has a very clumpy distribution 
not only in the plane of the sky but also along the line of sight. The \2 emission shown in the three first panels 
in Fig.\,\ref{12coReg2}-bottom may correspond to gas located in front (along the line of sight) of the SNR G18.9+0.3,  
while the emission displayed in the remaining panels would suggest a correspondence between the SNR and the molecular gas 
(see mainly the small clump towards the west at panel 24.5 and 26.5\ks). We note that the molecular emission towards 
the bottom left corner of the surveyed region, mainly at 22.5 \ks, should belong to the upper right border of the structure 
analysed in Reg.1. Thus, even though the gas distribution in the last panels of Fig.\,\ref{12coReg2}-bottom may suggest a correspondence between
the SNR and the molecular gas, 
these results allow us to conclude that the unusual flattened border of SNR G18.8+0.3 must have some physical origin other than interaction
with dense environmental gas.

\section{Summary and concluding remarks}

We report the analysis of the molecular gas towards the southern environment of the SNR G18.8+0.3 using ASTE observations. 
We have surveyed two regions, one towards  
the SNR southernmost `corner' which in addition harbors the \hii~region G018.630+0.309, and a second one towards the most flattened 
border of the SNR. The main results can
be summarized as follows: 

(a) In the first region, named Reg.1/Reg.1b, we discovered a molecular feature 
with a good morphological correspondence with the SNR southernmost corner, where the indentations in the morphology
of the SNR radio continuum emission are complemented by protrusions in the molecular cloud, strongly 
suggesting a SNR-molecular cloud interaction.
Additionally, analysing this molecular cloud with the \3 J=3--2 line we found a molecular feature 
composed by two clumps, not correlated with the \2 peak, that partially surrounds the \hii~region G018.630+0.309 and correlates with the
thermal dust emission at 870 $\mu$m. 
We found that the virial mass of the \2-clump considerably differ from the mass
value obtained through other methods, suggesting that it is 
not gravitationally bound. Finally, from a SED fitting to the IR counterpart of the \hii~region, we
confirm that it is located at the far distance, suggesting that it should be embedded in the molecular
cloud southwards the SNR.

(b) The second region, named Reg.2, covers a portion of the most flattened
border of SNR G18.8+0.3, where originally it was suggested an interaction with a dense cloud \citep{dubner99}. 
Strikingly, our \2 J=3--2 observations show a clumpy molecular structure without any morphological correspondence 
with the flat border of the SNR that could suggest
a causal relationship.
The better angular resolution attained in these new data served to
reveal that the flat southern border of the SNR must
originate in a different process other than the compression  of a
dense molecular cloud.

%Taking into account the results presented here and in Paper\,I, we conclude that SNR G18.8+0.3 is 
%encountering a large and clumpy molecular cloud that extends along its eastern and southern flanks. 
%We prove that the SNR is interacting with this cloud at its southern corner. On the other hand, we cannot 
%explain the reason of its most flattened south border in terms of an interaction with
%dense molecular material. 

%Additionally, we suggest that the progenitor of the SNR could be part of a 
%massive stars complex together with the exciting stars of at least 6 \hii~regions, four located in the Eastern Clump (see 
%Paper\,I) and two lying in the region studied in this work, which in turns are related to the same molecular cloud.

\section*{Acknowledgments}

{\tiny
The ASTE project is driven by Nobeyama Radio Observatory (NRO), a branch
of National Astronomical Observatory of Japan (NAOJ), in collaboration
with University of Chile, and Japanese institutes including University of
Tokyo, Nagoya University, Osaka Prefecture University, Ibaraki University,
Hokkaido University and Joetsu University of Education.
S.P., M.O., A.P., G.D., and E.G. are members of the {\sl Carrera del 
investigador cient\'\i fico} of CONICET, Argentina. M.C.P. is a doctoral
fellow of CONICET, Argentina. 
This work was partially supported by Argentina grants awarded by UBA (UBACyT), CONICET and ANPCYT.
M.R. wishes to acknowledge support from CONICYT through FONDECYT grant N$^{\rm o}$ 1140839.
A.P. is grateful to the ASTE staff for the support received during the observations. 
}

%%%%%%%%%%%%%%%%%%%%%%%%%%%%%%%%%%%%%%%%%%%%%%%%%%%%%%%%%%%%%%%%%%%%%
\bibliographystyle{aa}  % A&A format
   %\bibliographystyle{klunamed}     
   % format of references provided by the review (.bst)
\bibliography{biblio}
   % file containing the bibtex references (.bib)
\IfFileExists{\jobname.bbl}{}
{\typeout{}
\typeout{****************************************************}
\typeout{****************************************************}
\typeout{** Please run "bibtex \jobname" to optain}
\typeout{** the bibliography and then re-run LaTeX}
\typeout{** twice to fix the references!}
\typeout{****************************************************}
\typeout{****************************************************}
\typeout{}
}

\label{lastpage}
\end{document}